\documentclass{article}
\usepackage[utf8]{inputenc}
\usepackage[english]{babel}

\usepackage{amsfonts}
\usepackage{amsmath}
\usepackage{amssymb}
\usepackage{makecell}
\usepackage{graphicx}
\usepackage{hyperref}
\usepackage{bbm}
\usepackage{comment}
\usepackage{authblk}
\usepackage{url}

\usepackage{natbib}
\newtheorem{condition}{Condition}

\newcommand{\E}{\mathbb{E}}

\newcommand{\N}{\text{N}}
\newcommand{\Cov}{\text{Cov}}

\usepackage{setspace}

\usepackage{xcolor}

\title{Going beyond accuracy: estimating homophily in social networks using predictions}
\author[1]{George Berry}
\author[2]{Antonio Sirianni}
\author[3]{Ingmar Weber}
\author[3]{Jisun An}
\author[1]{Michael Macy}
\affil[1]{Department of Sociology, Cornell University, \texttt{\{geb97, mwm14\}@cornell.edu}}
\affil[2]{Department of Sociology, Dartmouth College, \texttt{antonio.d.sirianni@dartmouth.edu}}
\affil[3]{Qatar Computing Research Institute \texttt{\{iweber, jan\}@hbku.edu.qa}}
\affil[ ]{This draft: \today}
\date{}

\begin{document}

\maketitle

\begin{abstract}

In online social networks, it is common to use predictions of node categories to estimate measures of homophily and other relational properties. However, online social network data often lacks basic demographic information about the nodes. Researchers must rely on predicted node attributes to estimate measures of homophily, but little is known about the validity of these measures. We show that estimating homophily in a network can be viewed as a dyadic prediction problem, and that homophily estimates are unbiased when dyad-level residuals sum to zero in the network. Node-level prediction models, such as the use of names to classify ethnicity or gender, do not generally have this property and can introduce large biases into homophily estimates. Bias occurs due to error autocorrelation along dyads. Importantly, node-level classification performance is not a reliable indicator of estimation accuracy for homophily. We compare estimation strategies that make predictions at the node and dyad levels, evaluating performance in different settings. We propose a novel ``ego-alter'' modeling approach that outperforms standard node and dyad classification strategies. While this paper focuses on homophily, results generalize to other relational measures which aggregate predictions along the dyads in a network. We conclude with suggestions for research designs to study homophily in online networks. Code for this paper is available at \href{https://github.com/georgeberry/autocorr}{\texttt{https://github.com/georgeberry/autocorr}}.

\end{abstract}

\section{Introduction}

Researchers have long sought to understand the pattern \citep{marsden_core_1987, mcpherson_social_2006}, causes \citep{kossinets_origins_2009}, and consequences \citep{blau_macrosociological_1977} of \emph{homophily} \citep{coleman_relational_1958}, or the tendency for like to associate with like. Measuring the similarity of nodes along along racial \citep{marsden_core_1987, mcpherson_social_2006, smith_social_2014, cesare_redrawing_2017, mollica_racial_2003, wimmer_beyond_2010}, ethnic \citep{hofstra_sources_2017}, gender \citep{messias_white_2017, thelwall_homophily_2009, choudhury_tie_2011}, social status \citep{kossinets_origins_2009}, cultural \citep{lewis_social_2012}, emotional \citep{himelboim_valence-based_2016}, political \citep{halberstam_homophily_2016, bakshy_exposure_2015, colleoni_echo_2014}, and socioeconomic \citep{diprete_segregation_2011} lines is a core area of research in network science \citep{mcpherson_birds_2001}. In online networks, understanding the structure of homophily is crucial for understanding echo chambers \citep{barbera_tweeting_2015}, access to information, and network integration \citep{karimi_homophily_2018}. 

\begin{figure}[h]
\centering
\includegraphics[width=\textwidth]{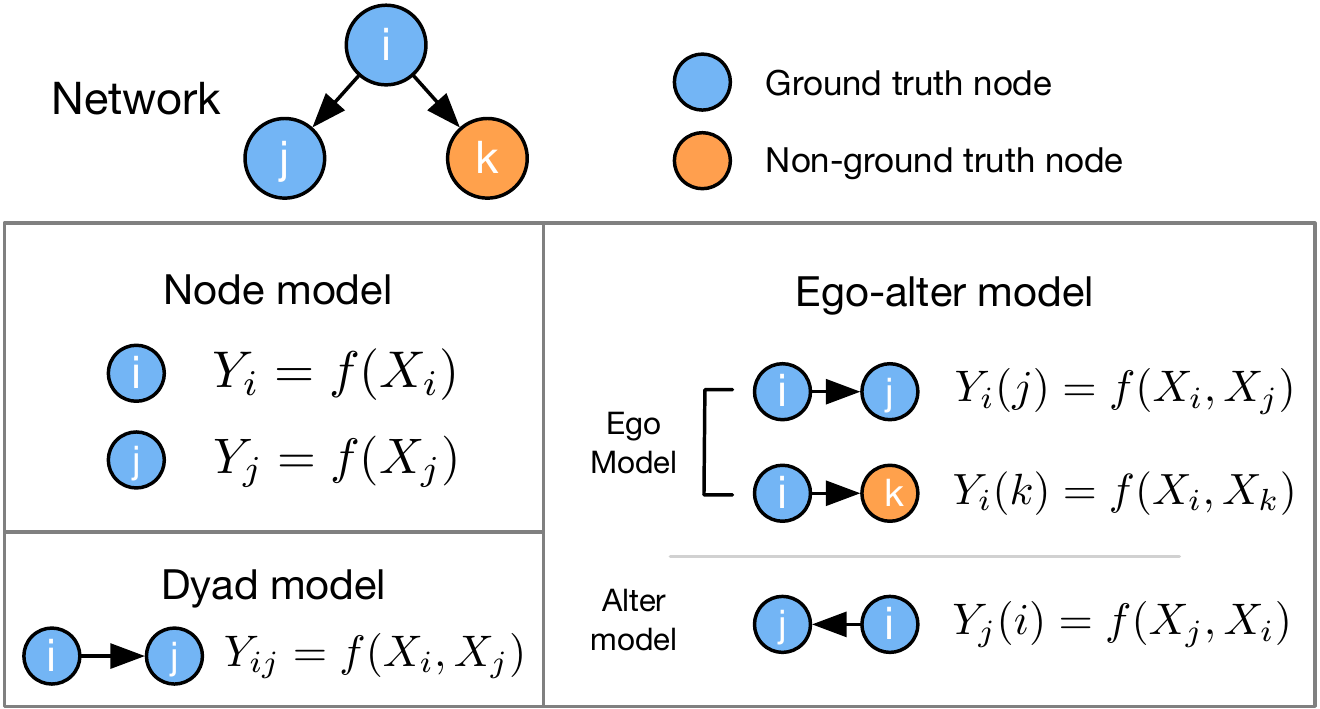}
\caption{A depiction of how the models studied in this paper (node, dyad, and ego-alter) turn a simple network into a prediction task. The node model trains on each ground truth node using only ego features $X_i$. The dyad model trains on each ground truth dyad using features from both ego and alter, $X_i, X_j$. The ego-alter model fits an ``ego model'' predicting ego's category for each link from ego, and a second ``alter model'' for each link incoming to each alter. Both ego and alter models incorporate features of ego and alter, $X_i$ and $X_j$, for each prediction, producing a ``family of predictions'' for each node.}
\label{fig:models}
\end{figure}

Online social networks present a particular challenge for understanding this fundamental aspect of networks: demographic and attitudinal information is often absent. The common strategy for addressing this is to predict demographics or attitudinal attributes \citep{cesare_redrawing_2017, messias_white_2017} based on publicly available information such as names, profile photos, text, or other information \citep{barbera_less_2016, al_zamal_homophily_2012, hofstra_predicting_2018, choudhury_tie_2011, messias_white_2017}. This information is combined with \emph{ground truth} labels (known values of the category of interest for a set of individuals), and a supervised learning classifier \citep{molina_machine_2019} is then used to predict the node category for all nodes in the network.

Although predicted node attributes are widely used to empirically measure homophily and other relational properties \citep{cesare_redrawing_2017, messias_white_2017, himelboim_valence-based_2016, boutyline_social_2017, hobbs_online_2016, bakshy_exposure_2015, colleoni_echo_2014,choudhury_tie_2011}, there is a lack of theoretical methodological work investigating when and to what extent the predictions produce reasonable estimates \citep{berry_estimating_2018}. The most common strategy is to choose a model which maximizes a node-level measure of classification performance. Because of the complexities of networks, this criterion is not sufficient for reliable estimation of relational measures such as homophily.

In this paper, we formalize the homophily estimation problem as a dyadic prediction problem. This expression clarifies the difficulty in using node-level predictions to draw larger inferences: node residuals are multiplied along edges, magnifying a node's residual in proportion to its degree, and opening the door to residual autocorrelation along dyads. Theoretically, we should expect correlated errors along edges due to latent homophily \citep{dellaposta_why_2015}, or the correlation of unobserved factors in the network. For example, name-based classifiers \citep{hofstra_predicting_2018, choudhury_tie_2011, hobbs_online_2016} have frequently been used for gender classification. If a name-based method codes ``Leslie'' as ``woman'' because this is more common in the population, yet for a specific network community the name ``Leslie'' tends to indicate ``man'', model errors will be correlated with the network and can bias overall gender homophily estimates. This issue is compounded by the highly skewed degree distributions of online networks \citep{kato_network_2012}, which introduces the additional possibility that misclassification for high degree nodes will bias the overall estimate.

We show that dyad-level predictions produce unbiased homophily estimates. However, such estimates are often high-variance for a given ground truth labeling budget\footnote{This fact suggests that even when possessing the ``ideal'' random edge sample with labels, modeling should be employed as a variance reduction technique.}. This motivates a two-step modeling procedure (called ``ego-alter'') which predicts the category of a node from the perspective of each one of its network neighbors. This allows incorporating network information beyond the ego level, while still using standard modeling tools such as logistic regression. This ego-alter approach is theoretically less biased than a node-level model, and across a range of simulations outperforms both node-level and dyad-level models in overall error. Figure \ref{fig:models} visually compares these three approaches. While we primarily study homophily with node demographics in mind, results here apply to a wide range of networked outcomes, such as estimating the fraction of people belonging to a certain race/ethnicity experience hate speech in their social media feeds \citep{davidson_racial_2019}.

\begin{figure}[h]

\includegraphics[width=0.8\textwidth]{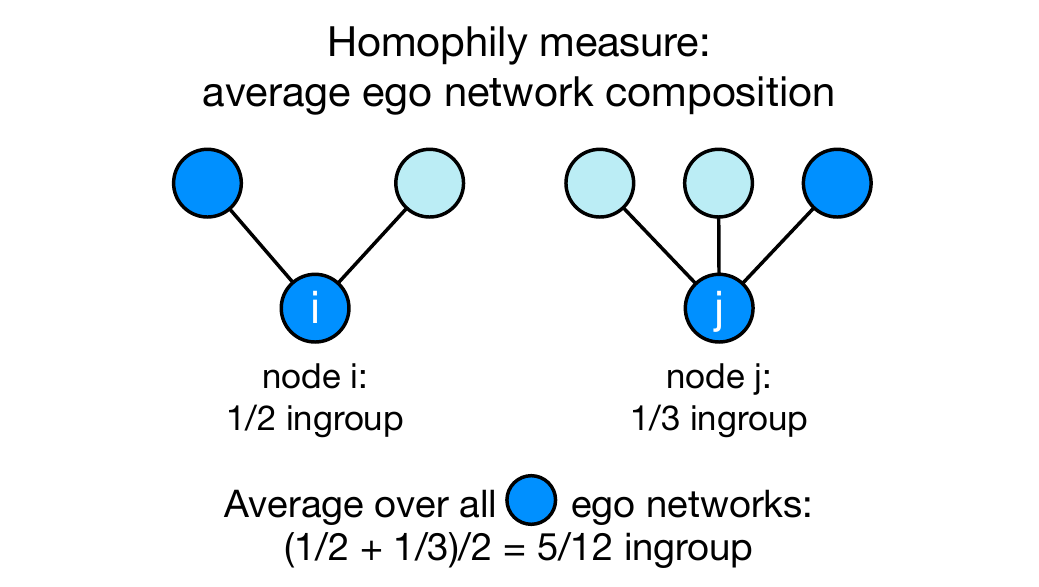}
\centering 
\caption{We use average ego network composition to capture homophily from the perspective of the dark blue nodes. We assume node categories (light and dark blue) must be predicted with a model. This estimand is expressed analytically in Equation \ref{eq:estimand_edge}.}
\label{fig:estimand}
\end{figure}

\section{Homophily Measure} \label{sec:measure}

We study the average fraction of ego networks composed of ingroup members (visualized in Figure \ref{fig:estimand}). Average ego network composition has been extensively studied in sociology, primarily in research concerning the General Social Survey network module \citep{marsden_core_1987, mcpherson_social_2006}. The average ego network composition measures what the network tends to look like from the perspective of members of a given group. For instance, Black respondents to the GSS have been found to have higher average racial heterogeneity in their core discussion networks than White respondents\footnote{We choose this measure of homophily instead of Coleman's homophily measure \citep{coleman_relational_1958} because it is not dominated by high degree nodes, although the Appendix shows that results for the average egonet measure apply to Coleman's measure as well.} \citep{marsden_core_1987}.

Average egonet composition can be written as a sum over ego networks, taking into account the category of both ego and alter. Let $Y_i$ indicate the category of $i$, for instance in the case of racial homophily, category $a$ may indicate Black, category $b$ may indicate White, and so on. Without loss of generality, assume that we are studying a binary outcome where $Y_i \in \{a, b\}$. For compactness, we write $Y_i^a$ to denote $\mathbbm{1}_{Y_i = a}$. Then the average fraction of group $a$'s ego networks which are composed of alters in group $a$ (Figure \ref{fig:estimand}) can be written,

\begin{equation} \label{eq:raw}
    H^{aa} = T[Y_i^a]^{-1} \sum_i Y_i^a \frac{1}{D_i} \sum_{j \in \N(i)} Y_j^a,
\end{equation}
where $D_i$ indicates the degree of node $i$, $\N(i)$ is a function which returns the neighbors of $i$, and $T[Y_i^a]$ indicates the size of group $a$, $\sum_{i \in V} Y_i^a$. For example, if $H^{aa} = 0.7$, it means that an average ego network for group $a$ is composed of 70\% ingroup members.

Note that equation \ref{eq:raw} can be re-written as a sum over dyads in the network by rearranging the summation,

\begin{equation} \label{eq:estimand_edge}
    H^{aa} = T[Y_i^a]^{-1} \sum_{(i, j) \in E} \frac{1}{D_i} Y_i^a Y_j^a =  T[Y_i^a]^{-1} \sum_{(i, j) \in E} \frac{1}{D_i} Y_{ij}^{aa},
\end{equation}
where $E$ are edges in graph $G$. Rewriting the edge-level outcome $Y_i^a Y_j^a$ as a single random variable $Y_{ij}^{aa}$ provides an expression in terms of edge categories. Estimating $H^{aa}$ can therefore be considered either a node or dyadic prediction task. Note that in the node case, predictions are multiplied.

\section{Dyadic regression as an unbiased estimator} \label{sec:dyadic}

Equation \ref{eq:estimand_edge} shows how homophily can be estimated with knowledge of edge categories $Y_{ij}^{aa}$. Assume edge categories $Y_{ij}^{aa}$ are obtained for a random sample of the edges $S$, and features correlated with edge categories $X_{ij}$ are available both for the sample $S$ and the population $P$. Assuming random sampling simplifies the argument, and we discuss deviations from this assumption in the Appendix.

Assume a model predicting $Y_{ij}^{aa}$ is chosen which has the property that the residuals sum to zero in the population and are uncorrelated with features\footnote{For instance, ordinary least squares and logistic regression both have this property, as does any model with mean squared error loss.} $\sum_{ij} e_{ij} = 0$, $\Cov(e_{ij}, X_{ij}) = 0$. Then this model trained on the sample $S$ provides an unbiased estimate of homophily in the population given that $\frac{1}{D_i}$ is included in $X_{ij}$ as a feature.

The reason for including $\frac{1}{D_i}$ as a variable can be seen by the following argument. First, recall the conditional expectation (CEF) function expression \citep{angrist_mostly_2009}: $Y_{ij}^{aa} = \E[Y_{ij}^{aa} | X_{ij}] + e_{ij}$, where $\E[Y_{ij}^{aa} | X_{ij}]$ can be estimated with a model such as OLS. An estimator for Equation \ref{eq:estimand_edge} can be written in terms of predictions,

\begin{equation} \label{eq:estimand_edge_hat}
    \hat{H}^{aa} = T[Y_i^a]^{-1} \sum_{(i, j) \in E} \frac{1}{D_i} \E[Y_{ij}^{aa} | X_{ij}].
\end{equation}
We now need to examine when using model predictions in Equation \ref{eq:estimand_edge_hat} provides an answer equal to Equation \ref{eq:estimand_edge} in expectation. This can be done by substituting the CEF into Equation \ref{eq:estimand_edge_hat} to obtain,

\begin{equation} \label{eq:bias}
    \hat{H}^{aa} =
    \underbrace{
        T[Y_i^a]^{-1} \sum_{(i, j) \in E} \frac{1}{D_i} Y_{ij}^{aa}
    }_{\text{True value}} -
    \underbrace{
        T[Y_i^a]^{-1} \sum_{(i, j) \in E} \frac{e_{ij}}{D_i}
    }_{\text{Sum of residuals}}.
\end{equation}
This indicates that the homophily estimate will be unbiased when the sum of residuals term is zero.

\begin{condition} \label{cond:unbiased}
When $\E[\sum_{(i, j) \in E} \frac{e_{ij}}{D_i}] = 0$, the expectation of the estimate equals the true value, $E[\hat{H}^{aa}] = H^{aa}$.
\end{condition}
Since we assumed a model is used where $\sum_{ij} e_{ij} = 0$ and $\Cov(e_{ij}, X_{ij}) = 0$, if $\frac{1}{D_i}$ is included in $X_{ij}$ then $\E[\sum_{ij} \frac{1}{D_i} e_{ij}] = 0$.


This argument concerns model \emph{residuals}, not error terms. No assumptions have been made about causality, true functional form, or predictive accuracy. With a random edge sample and OLS, $\frac{1}{D_i}$ is the only required variable in for an unbiased estimate, although this can produce a high variance estimate. Including robust predictive features is therefore still important for variance reduction and address cases of non-random sampling.

\section{Approximating dyadic regression with node-level data} \label{sec:ego_alter}

Sampling and labeling limitations often make collecting a large set of ground-truth dyads infeasible. In this situation, node-level ground truth information can be employed to estimate homophily. We present a two-step modeling strategy which we term ``ego-alter'' which uses only node-level ground truth information, reduces bias over a standard node-level model, and reduces variance compared to the edge model in Section \ref{sec:dyadic}. The ego-alter approach is biased, although the magnitude of bias in simulations we examine below is generally substantially less than a node-level model.

The ego-alter model is a hybrid approach: it uses dyadic features $X_{ij}$ to predict the node-level ground truth $Y_i$ and $Y_j$ separately, producing one prediction per edge for both ego and alter (see Figure \ref{fig:models} for a visual representation). This has the benefit of reducing bias in two ways: first, predictions for $Y_i$ and $Y_j$ are improved by including a richer set of features which improves prediction accuracy; second, it reduces bias by reducing dyadic residual autocorrelation.

Since $Y_{ij}^{aa} = Y_i^a Y_j^a$, $H^{aa}$ can be estimated with the product of node predictions,
\begin{equation*}
\begin{split}
    \hat{H}^{aa} &=  T[Y_i^a]^{-1} \sum_{(i, j) \in E} \frac{1}{D_i} E[Y_i^a | X_{ij}] E[Y_j^a | X_{ij}] \\
    & = T[Y_i^a]^{-1} \sum_{(i, j) \in E} \frac{1}{D_i} (Y_i^a - e_i^a) (Y_j^a - e_j^a),
\end{split}
\end{equation*}
where the second line follows by substituting the CEF. This can be expressed as the true homophily value plus two bias terms,
\begin{equation} \label{eq:equality_full}
    \hat{H}^{aa} =
    \underbrace{
        \frac{\sum_{(i, j)} \frac{1}{D_i} Y_i^a Y_j^a}{T[Y_i^a]}
    }_{\text{True value}} -
    \underbrace{
        \frac{\sum_{(i, j)} \frac{1}{D_i} e_i^a Y_j^a}{T[Y_i^a]}
    }_{R_1} -
    \underbrace{
        \frac{\sum_{(i, j)} \frac{1}{D_i} E[Y_i^a | X_{ij}] e_j^a }{T[Y_i^a]}
    }_{R_2}.
\end{equation}
The bias terms $R_1$ and $R_2$ both indicate dyadic correlation of the model residuals with neighbor outcomes. Assuming $\frac{1}{D_i}$ is included as a model feature, $R_1$ and $R_2$ can be thought of similarly: when model residuals are correlated with neighbor outcomes, the terms will be non-zero. This can happen when models produce pockets of similar errors in the network due to unobserved, network correlated features. When inverse ego degree $\frac{1}{D_i}$ is not included as a model feature, the bias terms become substantially more complex because of the interaction between degree, errors by degree, and errors along dyads.

Equation \ref{eq:equality_full} indicates that the estimate equals its true value when $R_1 = R_2 = 0$ or when $R_1 = -R_2$. Note that $R_1 = -R_2$ is unlikely, this is because residuals for $e_i$ and $e_j$ have similar correlations with neighbor true outcomes $Y_j$ and $Y_i$ since both ego and alter models score the entire network\footnote{This is confirmed by simulations, where $R_1$ and $R_2$ tend to have similar values.}.

Note that $R_2$ is the result of combining two terms, since $E[Y_i^a | X_{ij}] e_j^a = Y_i^a e_j^a + e_i^a e_j^a$. This suggests an ``augmented'' ego-alter model: first, fit an ego model for $i$, and then fit an alter model for $j$ which includes the ego predictions $E[Y_i^a | X_{ij}]$ as a feature. This, in expectation, eliminates the $R_2$ term and reduces bias assuming $R_1$ and $R_2$ have the same signs.

\section{Simulation} \label{sec:simulation}

We use simulations to evaluate the effectiveness of dyadic regression and ego-alter regression for estimating average egonet composition (Equation \ref{eq:estimand_edge}; Figure \ref{fig:estimand}). For an outcome $Y_i$ which takes on values $\{a, b\}$, the probability of $Y_i = a$ is simulated as follows:
\begin{equation}
    Y_i^a \sim \text{Bernoulli}(p_i), \; p_i = \text{logit}(2 X_i + Z_i),
\end{equation}
where $X_i$ and $Z_i$ represent individual and network-correlated features, respectively. The individual-level feature is drawn from a normal distribution, $X_i \sim \mathcal{N}(0, 1)$, while the network feature is the maximum of the individual feature among the neighbors of each node $i$: $Z_i = \max_{j \in \text{N}(i)}(X_j)$. $Z_i$ is then standardized to follow a normal distribution. This creates outcomes correlated along some dyads in the network, where nodes with large values of $X_i$ ``influence'' neighbors. If $Z_i$ is omitted, model residuals will be correlated along dyads and bias homophily estimates. The level of homophily generated by these parameters is moderate: the average ego network for group $a$ contains 59\% of nodes in group $a$ ($H^{aa} = 0.59$), while Coleman's homophily index for group $a$ is 0.14.

We choose $Z_i$ to be the maximum $X_j$ value among the alters $j$ of ego $i$ to provide a challenging setting for models: the true response is determined at the ego-network level yet models operate at the node or dyad levels, meaning a dyadic regression cannot capture the true functional form of the data generating process. This both approximates real-world scenarios where the data generating process is unknown, and demonstrates the argument about bias in Section \ref{sec:dyadic}. Five alternative simulation specifications are examined in the Appendix, with qualitatively similar results to this simulation.

Networks with 4000 nodes are generated according to a preferential attachment graph \citep{barabasi_emergence_1999} with five links per node and a powerlaw exponent $k=0.8$. Links are considered bidirected. Preferential attachment graphs have high degree disparities, providing a challenging setting for the estimation task considered here, since model errors on individual high degree nodes can bias estimates. We conduct simulations for both node and edge sampling, selecting 20\% of nodes or 2.5\% of edges randomly as ground truth cases. This produces roughly 800 ground truth nodes for both types of sampling. Note that sampling nodes into the ground truth set provides some ground truth dyads (and vice versa), meaning both node and dyad models can be fit with either type of sampling.

Using the ground truth sample to estimate a model, we classify all edges and estimate homophily across 500 simulation runs. Model performance is estimated in two ways: bias and absolute error. Bias is the average of $(\hat{H}^{aa} - H^{aa}) / H^{aa}$ across all simulation runs, and represents the systematic deviation from the true value. Absolute error is the average absolute error relative to the true underlying value, or the average of $|\hat{H}^{aa} - H^{aa}| / H^{aa}$ across all simulation runs. It captures how far estimates tend to be from the true value. 

Since both absolute error and bias are normalized, they have the interpretation of ``percent error.'' The bias-variance tradeoff means that we should not expect the method with the lowest bias to also have the lowest absolute error.

We evaluate three types of models: node, dyad, and ego-alter (see Figure \ref{fig:models} for a depiction), all fit with logistic regression. For the node model, we examine models with and without network features. For the ego-alter model, we examine both the basic version and the ``augmented'' version. This gives a total of 5 models, which are described here in terms of their regression equations, where $f(X_i, X_j)$ indicates a main-effects linear model $\beta_0 + \beta_1 X_i + \beta_2 X_j$.

\begin{equation*} 
\begin{split}
     \text{Node (no network)}    \;\;\; & Y_i^a = f(X_i) + e_i \\
     \text{Node}                 \;\;\; & Y_i^a = f(X_i, \frac{1}{D_i}, D_i) + e_i \\
     \text{Dyad}                 \;\;\; &
        Y_{ij}^{aa} = f(X_i, X_j, \frac{1}{D_i}, D_i, D_j) + e_{ij} \\
     \text{Ego-alter}            \;\;\; &
        Y_i^a(j) = f(X_i, X_j, \frac{1}{D_i}, D_i, D_j) + e_{i(j)} \\
                                 \;\;\; &
        Y_j^a(i) = f(X_i, X_j, \frac{1}{D_i}, D_i, D_j) + e_{j(i)} \\
     \text{Ego-alter (augmented)} \;\;\; &
        Y_i^a(j) = f(X_i, X_j, \frac{1}{D_i}, D_i, D_j) + e_{i(j)} \\
                                 \;\;\; &
        Y_j^a(i) = f(X_i, X_j, \frac{1}{D_i}, D_i, D_j, \hat{Y}_i^a(j)) + e_{j(i)} \\
\end{split}
\end{equation*}
The notation $Y_i^a(j)$ indicates that we predict $i$'s category for each neighbor $j$ separately, using features of both $i$ and $j$ in the prediction. Ego and alter degree are included in models because they tend to reduce the bias and variance of estimates and are available to researchers conducting network studies.

\subsection{Simulation results}

\begin{figure}[h]
\centering
\includegraphics[width=\textwidth]{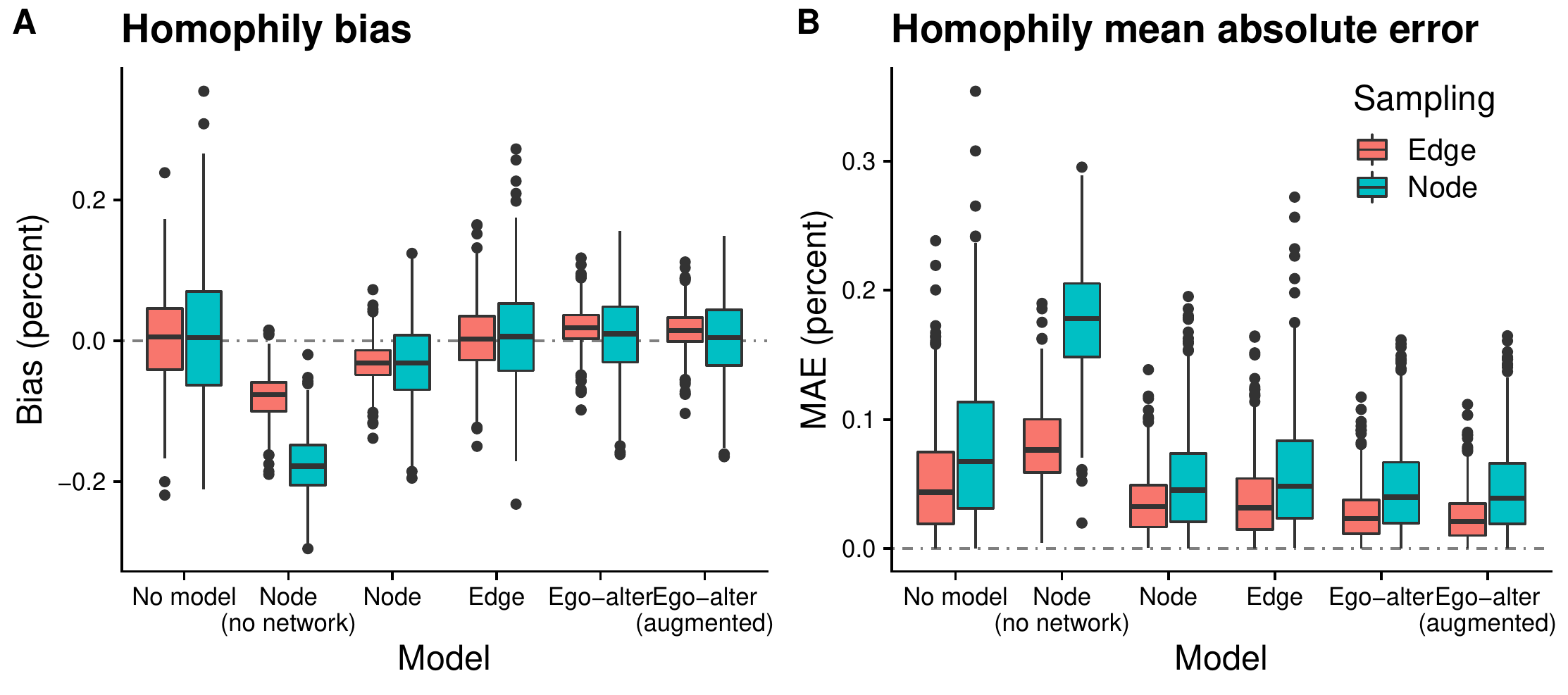}
\caption{The bias and absolute error of homophily estimates using five different models, for random node and random edge sampling. Node level models without network variables display large biases in the presence of network-correlated unobserved features. Including network information reduces this bias, and using edge or ego-alter models reduces this bias further. Note that while dyadic regression is unbiased, it does not provide the lowest error estimates. Since roughly similar numbers of nodes are sampled in both edge and node sampling, edge sampling is more efficient.}
\label{fig:results}
\end{figure}

As shown in Figure \ref{fig:results}, the default approach of using node-level classifier with no network features performs poorly. Homophily is underestimated by between 10\% and 20\%, with average absolute error of about the same magnitude. Even when accounting for the inverse degree term $\frac{1}{D_i}$, the node-level approach still underestimates homophily by around 3\%. This large reduction in bias indicates the importance of including network information in the model predicting node categories, while the remaining bias indicates the limitations of a node-level approach in the presence of network-correlated outcomes.

In this simulation, homophily is under-estimated. This indicates that errors tend to be positively correlated along dyads, increasing the sum of residuals term in Equation \ref{eq:bias} and reducing the overall homophily estimate. In other words, there are pockets of the network where the model errors are similar. An alternative scenario exists where a model produces negatively correlated dyadic errors and an over-estimate of homophily. An instance where this happens is residual-degree correlation in the network. When high degree nodes have positive residuals and low degree nodes have negative residuals, the overall residual term in Equation \ref{eq:bias} can be negative and cause an over-estimate of homophily\footnote{An instance of this phenomenon can be seen in the Section \ref{sec:five_sims}, with the simulation called ``degree.''}.

A dyadic model produces an unbiased estimate of homophily, according with the argument in Section \ref{sec:dyadic}. However, the dyadic approach does not produce the lowest absolute error. Despite a small amount of bias (around 1\%), the ego-alter approach produces lower absolute error on average than the dyadic approach. In alignment with the theoretical argument in Section \ref{sec:ego_alter}, including ego predictions in the alter model reduces bias about 20\% on average.

While this simulation uses random sampling, the ego-alter model is also more robust to deviations from random sampling compared to other methods, as shown in the Appendix. In the presence of non-random edge samples, an edge model can be brittle. One potential corrective is weighting the ground truth data, but the often high-dimensional nature of edge features risks large design effects due to the curse of dimensionality \citep{iacus_causal_2012}. Additionally, a ``meta-network-correlation'' problem can arise, where errors in weights are network correlated.

A more practical approach is to employ a modeling strategy such as ego-alter which can more flexibly learn class probabilities in a network-aware way. While we intentionally restrict models here to logistic regression with only main effects, more flexible functional forms can also be employed to better approximate class probabilities within subgroups.

\subsection{Node-level performance and network-level estimands}

Machine learning models are usually evaluated on observation-level performance metrics such as precision, recall, and area under the curve (AUC). When using predictions to estimate an aggregate such as homophily, strong observation-level performance is encouraging but not sufficient for high-quality estimates of the aggregate. An error-free model will by definition produce a perfect estimate of homophily, but even models with strong out of sample observation-level performance can make dyad-autocorrelated errors that bias homophily estimates.

\begin{figure}[h]
\centering
\includegraphics[width=\textwidth]{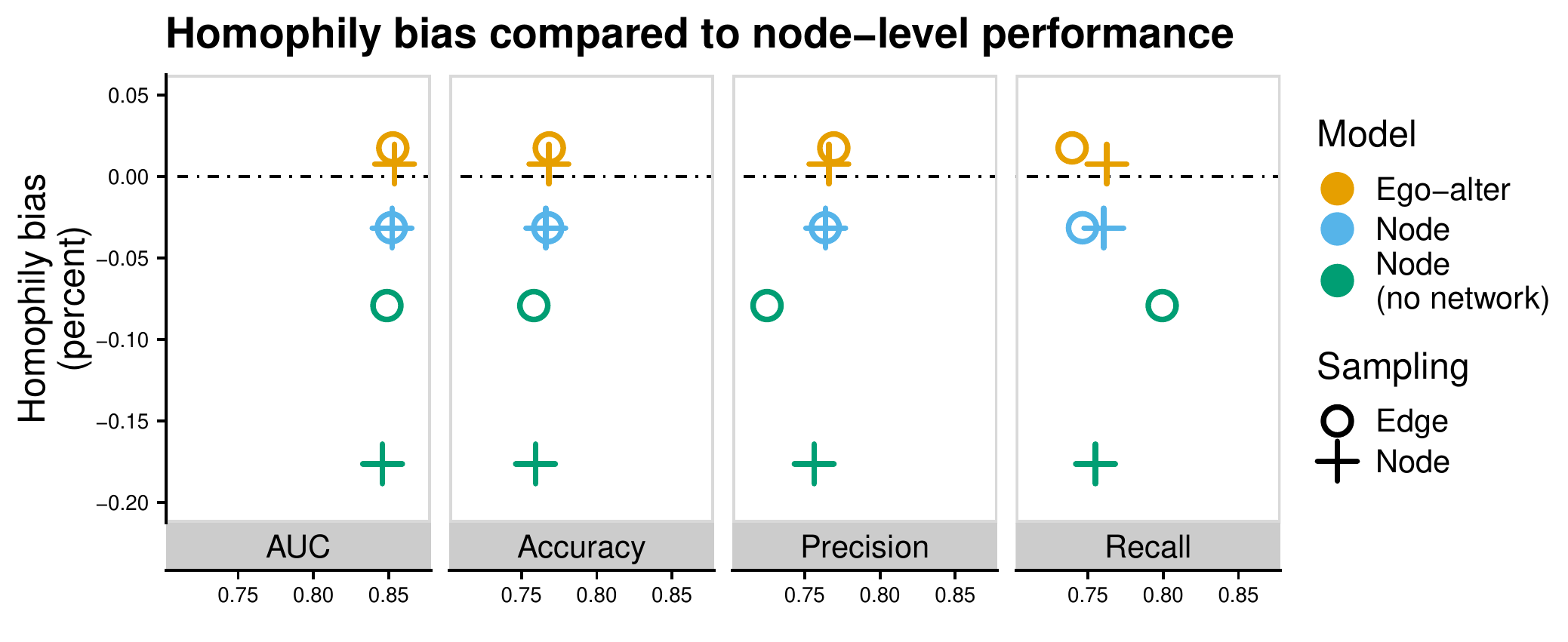}
\caption{Models with similar node-level classification performance produce different levels of bias when estimating homophily. The three models in Figure 4 have average AUCs between 0.846 and 0.853, yet produce average biases ranging from 0.8\% to 17.6\%. This demonstrates that observation-level classification performance and estimation of relational measures are distinct tasks.}
\label{fig:traditional}
\end{figure}

This can be seen clearly in Figure \ref{fig:traditional}, which plots model performance against bias in estimating homophily\footnote{Only models which produce node-level category predictions can be evaluated this way, which necessitates excluding the edge model.}. Models differ only slightly on traditional performance measures, yet produce large differences in homophily bias. The best model's AUC is 0.8\% better than the worst model, yet has a bias reduction of 95\% (worst: 17.6\% bias; best 0.8\% bias).

Note that a meta-analysis of research on demographic classification on social media \citep{cesare_how_2017} found a median accuracy of 0.81 for predicting race/ethnicity, while simulations presented here have an average accuracy of around 0.77. This indicates that similar biases may be present with the type of classification performance found in real-world tasks.

\section{Practical advice} \label{sec:advice}

When studying homophily in online communities, researchers can potentially improve the quality of estimates in five ways: including network information in models, using the ego-alter modeling strategy, improving model flexibility, sampling edges, and using cross-validation to check for the presence of network-residual correlation.

First and most importantly, network information should be incorporated into prediction models. Evidence from Sections \ref{sec:simulation} and \ref{sec:five_sims} indicates the single largest improvement in model performance comes from including degree information ($\frac{1}{D_i}$) in models. The specific information to include is dependent on the estimand, and can even extend to behavioral information when outcomes such as political affiliation are studied in networks. We give an example of applying the process from Section \ref{sec:dyadic} to new estimands in the Appendix (Sections \ref{sec:other_factors} and \ref{sec:coleman}).

Second, the simulation results consistently demonstrate that the ego-alter modeling strategy performs well both in terms of bias and absolute error. This is true even in the presence of a non-random ground truth sample. Since the ego-alter strategy is new, we recommend that researchers present results from both a node-level model and an ego-alter model, with network information included for both models. 

Third, and closely related to the choice of modeling strategy is the choice of model itself: a logistic regression with only main effects in the presence of a non-random sample can produce large biases, as seen with the edge model and non-random edge sampling in the Appendix (Table \ref{tab:bias_sims}). A more flexible model can mitigate this by better learning conditional class probabilities, although the performance will depend on having sufficient ground truth data.

Fourth, edges should be sampled instead of nodes when possible.  A consistent finding of our simulations is that for the same labeling budget, a random edge sample outperforms a random node sample in terms of absolute error. In practical settings, such as Twitter, it is often much easier to randomly sample nodes than edges. One strategy for edge sampling is to use a result from the respondent driven sampling literature \citep{salganik_sampling_2004} that a random walk through an undirected network approximates an edge sample (see \citep{berry_estimating_2018} for a discussion in the context of online networks). While this may not be feasible in some research settings, researchers may want to consider edge sampling if a random walk approach is possible.

Finally, researchers can obtain an estimate of network residual correlation by using cross-validation (see the discussion in \citep{molina_machine_2019} for a brief introduction to cross validation; see Chapter 7 of \citep{hastie_elements_2008} for a more extensive discussion). Cross validation splits the training data into a number of folds (usually 5 or 10), and uses all but one fold to train a model, with the held-out fold used to evaluate the model. This proceeds in a round-robin fashion so that the entire training set is scored in a way approximating out-of-sample prediction. In the context of homophily estimation, estimating the residual term in Equation \ref{eq:bias} can provide important information about network residual correlation. This can be accomplished in a cross validation setting by dividing up all dyads in the training set into folds, and performing cross validation on the ground truth dyads. If $\sum_{(i, j) \in S_{\text{train}}} \frac{1}{D_i} e_{ij} \neq 0$, where $S_{\text{train}}$ is the training set, then models may need adjustment before providing reliable estimates of homophily. This strategy does not ensure unbiased homophily estimates, particularly in the presence of non-random ground truth sampling, but it does provide a useful diagnostic.

\section{Conclusion}

We have examined the problem of estimating homophily when predictions must be used for node attributes. While the problem is challenging, the results we present indicate that homophily can be studied in online networks when classification performance is strong and network information is incorporated directly into models.

The strategies outlined here also provide a pathway for the measurement of other network-level properties. Examples are triadic properties, such as social closure by demographic group. In studies of dynamic network processes such as contagion, models to reduce measurement error \citep{berry_opacity_2019} may benefit from the results here. In the case of signed or multiplex networks, the distribution of different types of edges across groups may be important. Similarly to homophily estimation, consideration of how model errors intersect with graph properties is important for reliable use of predictions in networks.

\section{Appendix}

\subsection{Five simulations} \label{sec:five_sims}

In addition to the simulation presented in the main text, we examine four additional simulations. These simulations further demonstrate the strengths and weaknesses of the approaches considered in the main text. We describe these simulations by the data generating process for $Y_i^a$.

\begin{enumerate}
    \item \textbf{Independent}: $Y_i^a = 2 X_i$, where $X_i \sim \mathcal{N}(0, 1)$. In the independent simulation, node categories depend only on a node-level feature $X_i$ which is uncorrelated with the network.
    \item \textbf{Degree}: $Y_i^a = 2 X_i + Z_i$, where $X_i \sim \mathcal{N}(0, 1)$ and $Z_i = \text{mean}_{j \in \text{N}(i)}(X_j)$, $Z_i$ standardized. In this simulation, node categories are a function of neighbor degree, meaning nodes with high-degree neighbors are more likely to have $Y_i^a = 1$.
    \item \textbf{Main}: the simulation described in Section \ref{sec:simulation} of the main text.
    \item \textbf{Unobserved}: $Y_i^a = 2 X_i + Z_i$, where $X_i \sim \mathcal{N}(0, 1)$, $Z'_j \sim \mathcal{N}(0, 1)$ and $Z_i = \text{max}_{j \in \text{N}(i)}(Z'_j)$, $Z_i$ standardized. $Z'_j$ is a standard normal unobserved variable which causes outcomes to be correlated in the network.
    \item \textbf{Sampled}: This simulation samples nodes into the ground truth set according to degree and node features. $Y_i^a$ is generated identically to the main simulation. For the edge simulation, dyads are sampled into the ground truth set by, \[p((i, j) \in \text{ ground truth}) = \text{logit}(\alpha_{\text{dyad}} + 0.05 D_i + 0.05 D_j + 0.2 X_i + 0.2 X_j).\]
    
    For the node simulation, nodes are sampled into the ground truth set by, \[p(i \in \text{ ground truth}) = \text{logit}(\alpha_{\text{node}} + 0.05 D_i + 0.2 X_i).\]
    
    The $\alpha$ variables are constants chosen to make the total number of ground truth nodes or dyads equivalent to the sampling fractions for the other simulations.
\end{enumerate}

Tables \ref{tab:bias_sims} and \ref{tab:mae_sims} present all simulations by all models, including a ``no model'' category using just the ground truth observations for comparison. The best performing model for each simulation (each row) is bolded.

\begin{table}[ht]
\centering
\addtolength{\leftskip}{-2cm}
\addtolength{\rightskip}{-2cm}
\caption{Bias by simulation, with the best model in each row bolded.} 
\begin{tabular}{ll|rrrrrr}
  \hline
	\thead{Sampling\\level} & Simulation & No model & 	\thead{Node\\(no network)} & Node & Edge & Ego-alter & 	\thead{Ego-alter\\(augmented)} \\ 
  \hline
Edge & Independent & 0.003 & 0.001 & \textbf{-0.0003} & 0.001 & -0.0005 & -0.001 \\ 
       & Degree & 0.005 & 0.089 & 0.058 & \textbf{0.006} & 0.019 & 0.019 \\ 
       & Unobserved & 0.001 & -0.053 & -0.006 & \textbf{-0.001} & 0.018 & 0.018 \\ 
       & Main & 0.002 & -0.079 & -0.031 & \textbf{0.003} & 0.018 & 0.014 \\ 
       & Sampled & -0.947 & -0.064 & -0.031 & 0.623 & 0.025 & \textbf{0.021} \\ 
  Node & Independent & 0.004 & -0.004 & 0.001 & 0.001 & 0.0003 & \textbf{0.0002} \\ 
       & Degree & 0.006 & 0.114 & 0.033 & 0.002 & \textbf{0.001} & \textbf{0.001} \\ 
       & Unobserved & 0.004 & -0.145 & -0.003 & 0.005 & \textbf{-0.001} & \textbf{-0.001} \\ 
       & Main & 0.007 & -0.176 & -0.032 & 0.006 & 0.008 & \textbf{0.003} \\ 
       & Sampled & -0.057 & -0.116 & -0.032 & 0.030 & 0.015 & \textbf{0.011} \\ 
   \hline
\end{tabular}
\label{tab:bias_sims}
\end{table}

\begin{table}[ht]
\centering
\addtolength{\leftskip}{-2cm}
\addtolength{\rightskip}{-2cm}
\caption{Mean absolute error by simulation, with the best model in each row bolded.} 
\begin{tabular}{ll|rrrrrr}
  \hline
	\thead{Sampling\\level} & Simulation & No model & 	\thead{Node\\(no network)} & Node & Edge & Ego-alter & 	\thead{Ego-alter\\(augmented)} \\ 
  \hline
Edge & Independent & 0.054 & 0.030 & 0.024 & 0.045 & \textbf{0.023} & \textbf{0.023} \\ 
       & Degree & 0.063 & 0.090 & 0.060 & 0.051 & \textbf{0.030} & \textbf{0.030} \\ 
       & Unobserved & 0.051 & 0.055 & \textbf{0.026} & 0.042 & 0.029 & 0.029 \\ 
       & Main & 0.051 & 0.079 & 0.035 & 0.038 & 0.027 & \textbf{0.025} \\ 
       & Sampled & 0.947 & 0.081 & 0.058 & 1.585 & 0.051 & \textbf{0.049} \\ 
  Node & Independent & 0.093 & 0.044 & 0.060 & 0.076 & \textbf{0.059} & 0.060 \\ 
       & Degree & 0.107 & 0.116 & \textbf{0.072} & 0.088 & \textbf{0.072} & \textbf{0.072} \\ 
       & Unobserved & 0.076 & 0.145 & \textbf{0.049} & 0.069 & 0.050 & 0.050 \\ 
       & Main & 0.079 & 0.176 & 0.052 & 0.059 & 0.047 & \textbf{0.046} \\ 
       & Sampled & 0.060 & 0.116 & 0.037 & 0.040 & 0.029 & \textbf{0.028} \\ 
   \hline
\end{tabular}
\label{tab:mae_sims}
\end{table}

\subsection{Example extension: incorporating other factors into the estimand} \label{sec:other_factors}

Researchers often care about actions in addition to node characteristics. For instance, what is the fraction of content seen broken down by gender of the content author? Addressing this question is important for examining visibility \citep{karimi_visibility_2017} by gender online, and requires combining information about node gender and action.

In this case, Equation \ref{eq:estimand_edge} is modified with a variable $A_j$ indicating some action of alter $j$. Assume $A_j$ represents number of messages sent by $j$, and $Y_i = a$ indicates that $i$ is a woman.

\begin{equation} \label{eq:estimand_a}
    H^{aa}_{\text{extended}} = T[Y_i^a]^{-1} \sum_{(i, j) \in E} \frac{A_j}{D_i} Y_i^a Y_j^a.
\end{equation}
In words, Equation \ref{eq:estimand_a} represents the average fraction of messages seen by members of group $a$ which come from members of group $a$. When incorporating the additional variable $A_j$ into the equation, we can apply the same logic as Section \ref{sec:dyadic} to obtain an unbiased estimate: incorporate $\frac{A_j}{D_i}$ into the predictive model, instead of $\frac{1}{D_i}$ alone.

\subsection{Example extensions: Coleman's homophily index} \label{sec:coleman}

We studied average egonet composition in the main text, but another popular measure of homophily is Coleman's homophily index \citep{coleman_relational_1958}. This measure studies the fraction of within group links from the perspective of a certain group, relative to the proportion expected by chance.

The challenge is estimating the proportion of within-group links from the perspective of a given group $a$. This can be done in a manner similar to Equation \ref{eq:estimand_edge},

\begin{equation} \label{eq:coleman}
    H^{aa}_{\text{Coleman}} = T[Y_i^a]^{-1} \sum_{(i, j) \in E} Y_i^a Y_j^a.
\end{equation}
This turns out to be a simpler version of the egonet estimand considered in the main text, and can be addressed with similar modeling strategies.

\section{Acknowledgements}

We thank Thomas Davidson, Mario Molina, Pablo Barber\'a, Christopher Cameron, and the members of the Cornell Social Dynamics Lab for helpful comments and discussions.

\bibliography{dissbib}

\end{document}